\documentclass[conference]{IEEEtran}
\IEEEoverridecommandlockouts
\usepackage{cite}
\usepackage{amsmath,amssymb,amsfonts}
\usepackage{algorithmic}
\usepackage{graphicx,subcaption}
\usepackage{textcomp}
\usepackage{xcolor,newtxmath}

\usepackage{epsfig}
\usepackage{graphicx}
\usepackage{amsmath}
\usepackage{amssymb}
\usepackage{tikz}
\usepackage{enumitem}
\usepackage[font=small,skip=0pt]{caption}
\usepackage[font=small,skip=0pt]{subcaption}
\usepackage{setspace}
\usepackage{pdfpages}
\usepackage{fancyhdr}
\newbox{\bigpicturebox}
\def\etal{\textit{et al.}}
\newcommand{\Lagr}{\mathcal{L}}
\def\BibTeX{{\rm B\kern-.05em{\sc i\kern-.025em b}\kern-.08em
    T\kern-.1667em\lower.7ex\hbox{E}\kern-.125emX}}
\begin{document}

\title{Enhancing Perceptual Loss with Adversarial Feature Matching for Super-Resolution}
% {Augmenting Perceptual Loss with Adversarial Feature Matching for Super-Resolution}
% {Adversarial Enhancement of Perceptual Loss for Super-Resolution}
% {Enhancing Perceptual Loss Functions with Adversarial Feature Matching for Super-Resolution}
% An Improved Training Strategy for Super-Resolution

\author{\IEEEauthorblockN{Akella Ravi Tej$^\dagger$, Shirsendu Sukanta Halder$^{\text{*}\ddagger}$\thanks{* Equal contribution}, Arunav Pratap Shandeelya$^{\text{*}\S}$, Vinod Pankajakshan$^\dagger$}
\IEEEauthorblockA{$^\ddagger$ \textit{Carnegie Mellon University, USA} \\
$^\dagger$ \textit{Indian Institute of Technology Roorkee, India} \\
$^\S$ \textit{International Institute of Information Technology Bhubaneswar, India} \\
Email: ravitej.akella@gmail.com}
}

% \and
% \and
% \and
% \and
% \and

\maketitle

\begin{abstract}
Single image super-resolution (SISR) is an ill-posed problem with an indeterminate number of valid solutions. Solving this problem with neural networks would require access to extensive experience, either presented as a large training set over natural images or a condensed representation from another pre-trained network. Perceptual loss functions, which belong to the latter category, have achieved breakthrough success in SISR and several other computer vision tasks. While perceptual loss plays a central role in the generation of photo-realistic images, it also produces undesired pattern artifacts in the super-resolved outputs. In this paper, we show that the root cause of these pattern artifacts can be traced back to a mismatch between the pre-training objective of perceptual loss and the super-resolution objective. To address this issue, we propose to augment the existing perceptual loss formulation with a novel content loss function that uses the latent features of a discriminator network to filter the unwanted artifacts across several levels of adversarial similarity. Further, our modification has a stabilizing effect on non-convex optimization in adversarial training. The proposed approach offers notable gains in perceptual quality based on an extensive human evaluation study and a competent reconstruction fidelity when tested on objective evaluation metrics.
\end{abstract}


\thispagestyle{fancy}

\fancyhf{}

\renewcommand{\headrulewidth}{0pt}

\chead{\small Accepted for publication in the International Joint Conference on Neural Networks (IJCNN) 2020}

\pagestyle{empty}

\lfoot{\small \copyright 2020 IEEE. Personal use of this material is permitted. Permission from IEEE must be obtained for all other uses, in any current or future media, including reprinting/republishing this material for advertising or promotional purposes, creating new collective works, for resale or redistribution to servers or lists, or reuse of any copyrighted component of this work in other works.}

\section{Introduction}

%%%%%%% Figure

% \begin{figure*}[!t]
%     \begin{subfigure}[t]{0.45\textwidth}
%     \includegraphics{HR/HR_train.png}
%     \caption{HR}
%     \end{subfigure}
%     \begin{subfigure}[t]{0.45\textwidth}
%     \includegraphics{Other_Methods/Enhancenet/EnhaneNet_zoom.png}
%     \caption{EnhanceNet \cite{Enhancenet}}
%     \end{subfigure}\begin{subfigure}[t]{0.45\textwidth}
%     \includegraphics{Other_Methods/SRGAN/SRGAN_Zoom.png}
%     \caption{SRGAN \cite{SRGAN}}
%     \end{subfigure}\begin{subfigure}[t]{0.45\textwidth}
%     \includegraphics{OurMethod/MSE_VGG_PER_AT_Soft_Cov/Ours_zoom.png}
%     \caption{Bicubic}
%     \end{subfigure}

% \caption{Qualitative comparison using of \textit{Train} image from DIV2K on baseline models.}
% \label{fig:first}
% \end{figure*}

High-resolution ($HR$) images are perceived as more visually-pleasing than their corresponding mappings in low-resolution ($LR$) space since they form a better illusion of continuity.
% Upon increasing the resolution of an image, we are essentially capturing higher frequency contents that are responsible for shifting our perception of the image from jagged to continuous.
This perceived greater utility of $HR$ images over $LR$ images places a growing demand for signal processing techniques that learn a mapping between the $HR$ and $LR$ spaces. More formally, the problem of generating an $HR$ image from several $LR$ images is referred to as super-resolution reconstruction. 

% Super-resolution (SR) is a fundamental problem in computer vision with diverse applications such as 4K-HDTV's \cite{application_HDTV}, medical imaging \cite{application_medical_1}, etc.

\begin{figure}[!t]
    \centering
    \begin{subfigure}[t]{0.5\textwidth}
        \centering
        \includegraphics[height=2in,width=3in]{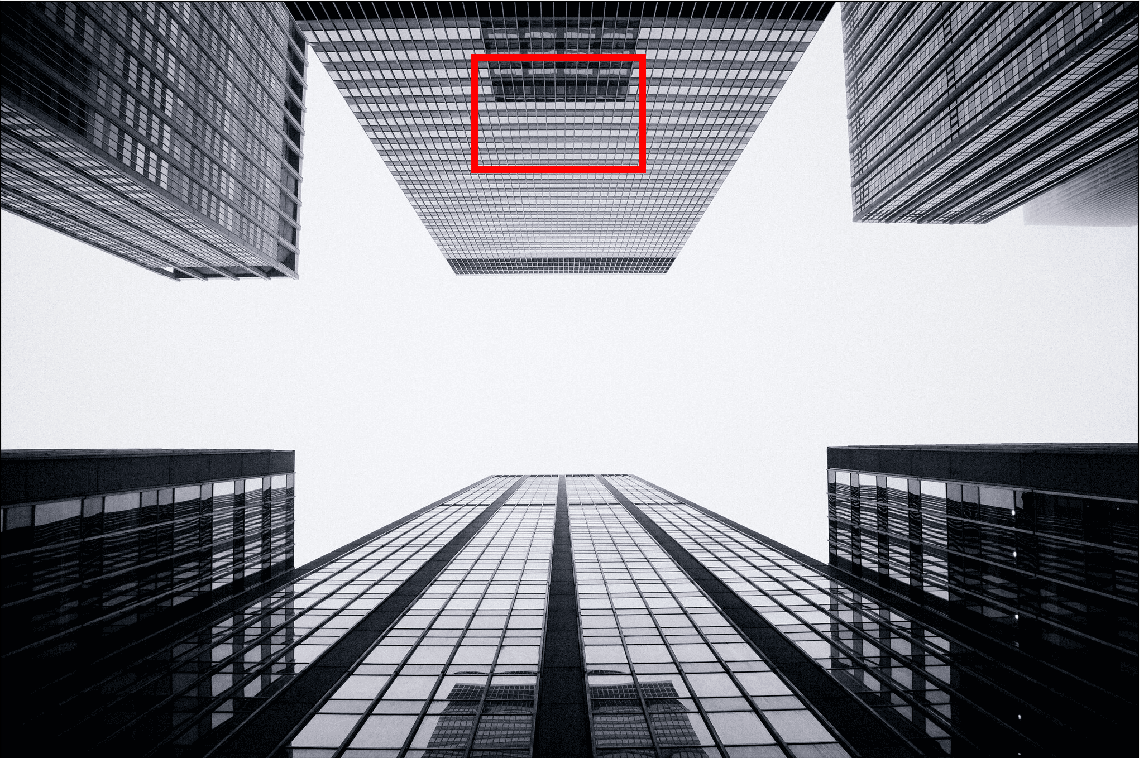}
        \caption{HR}
    \end{subfigure}
    \\
    \centering
    \begin{subfigure}[t]{0.23\textwidth}
        \centering
        \includegraphics[height=1in, width=1.5in]{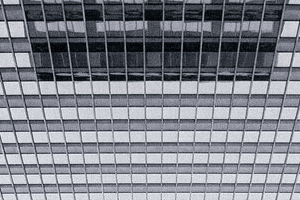}
        \caption{HR}
    \end{subfigure}
    \begin{subfigure}[t]{0.23\textwidth}
        \centering
        \includegraphics[height=1in, width=1.5in]{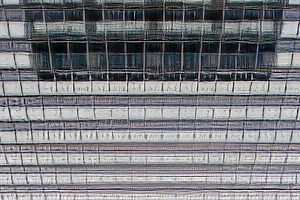}
        \caption{SRGAN \cite{SRGAN}}
    \end{subfigure}
    \begin{subfigure}[t]{0.23\textwidth}
        \centering
        \includegraphics[height=1in, width=1.5in]{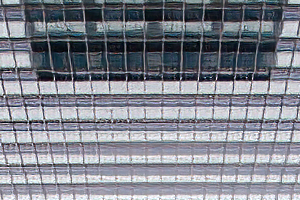}
        \caption{EnhanceNet \cite{Enhancenet}}
    \end{subfigure}
    \begin{subfigure}[t]{0.23\textwidth}
        \centering
        \includegraphics[height=1in, width=1.5in]{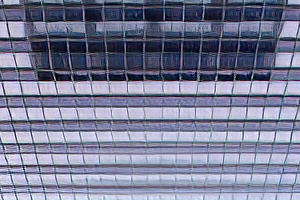}
        \caption{\textit{Ours ($M_{pc\sigma va}$)}}
    \end{subfigure}
\vspace{1em}
\caption{Demonstration of pattern artifacts introduced by perceptual loss. Comparing our method with state-of-the-art SR models that use perceptual loss on DIV2K (super-resolved images 
% from the algorithms 
are zoomed in
% on the portion of the red bounding box in the original HR image 
for better comparison).}
\label{fig:first}
\end{figure}

Our approach deals with a sub-problem of SR, where an $HR$ image needs to be reconstructed from a single $LR$ image, commonly known as \textit{Single Image Super-Resolution} (SISR). Since SISR is an ill-posed inverse problem with multiple valid $HR$ outputs for a single $LR$ input, modern supervised learning approaches \cite{SRCNN_1, VDSR, DRCN} restrict their solution space by learning a strong prior. For their capacity to learn complex and abstract representations, deep convolutional neural networks (CNNs) possess a favorable inductive bias for learning this prior. While CNNs trained on point-estimate loss functions attain state-of-the-art performance on peak signal-to-noise ratio (PSNR) metric, the generated images are overly smooth and have an unnatural appearance \cite{SRGAN}. Further, this problem worsens at high upscaling factors, causing a steep drop in the generation quality of SR images. To enforce photo-realism in the generated images, recent methods use perceptual loss \cite{JohnsonPerceptual, Enhancenet} and adversarial loss \cite{SRGAN} as objective functions for modeling the high-dimensional multi-modal distribution of natural $HR$ images. While generative adversarial networks (GANs) have shown great potential in the generation of visually-realistic images, their non-convex loss landscape results in unstable training. This severely limits current approaches that apply adversarial loss in combination with point-estimate loss or perceptual loss to ensure stable training. On the other hand, pre-trained perceptual loss functions provide a stable restoration of lost high-frequency components, although they often also introduce undesired artifacts in the generated outputs that current approaches fail to eliminate \cite{JohnsonPerceptual}.

% \begin{figure}[!htbp]
%     \centering
%     \includegraphics[width=0.9\linewidth]{mos.png}
%     \caption{MOS scores for combinations of different loss functions on our proposed model}
%     \label{fig:MOS_lossfunctions}
% \end{figure}

% \subsubsection{Qualitative and Quantitative Analysis}

%%%%%%% Figure

In this paper, we highlight that the pre-training objective of perceptual loss does not match with the true super-resolution objective. As a consequence, perceptual loss additionally transfers the biases from its pre-training stage that surfaces as pattern artifacts in the generated images. To address this objective mismatch, we propose a novel content loss
\footnote{Some works in the literature refer to content loss as feature matching loss.}
formulation that is an ensemble of content losses derived from the convolutional layers of a discriminator network. Each layer of the discriminator learns a unique abstraction for differentiating between the real and generated images, thereby allowing our content loss to address the removal of pattern artifacts that are identifiable across numerous levels of adversarial similarity. Further, we show that the proposed approach has connections with previously proposed techniques for stabilizing adversarial training in GANs. Finally, we conduct an extensive mean-opinion-score (MOS) test and the standard objective evaluation to demonstrate the gains in perceptual quality and content-preservation in the generated images.

\section{Background and Related Work}
\label{sec:prev_work}

\subsection{Single Image Super-Resolution (SISR)}

SISR involves the reconstruction of an $HR$ image while limiting the contextual information to a single $LR$ image.
% making it a more challenging problem when compared multi-image SR.
Early SISR approaches relied on filtering and interpolation \cite{bicubic}, generating overly smooth images.
% This behaviour is attributed to the ill-posed nature of SISR since there exist a large number of valid $HR$ interpolations for a given $LR$ image.
Example-based approaches address this issue by learning a strong prior from internal similarities in the same image \cite{Second_SISR} or by externally learning a mapping between the $LR$ and $HR$ patches. With sufficient data, external example-based approaches can be effectively implemented in standard supervised learning frameworks like sparse-representation coding \cite{peleg2014} and dictionary-based learning \cite{wang2012semi}.

% In recent times, deep CNN architectures have enjoyed a tremendous success in several vision-based applications such as image segmentation \cite{image_seg1}, object recognition \cite{obj_det1}, etc. This 
The recent success of deep CNN architectures
propelled Dong \etal \cite{SRCNN_1} into using a 3-layer CNN for SISR, which subsequently gave rise to a new direction of SR research with improved training methodologies. Kim \etal \cite{VDSR, DRCN} showed that recursive convolutions and residual learning could be used to realize deeper architectures that perform significantly better than shallow networks. To control the parameter count as we explore deeper architectures, Tai \etal \cite{DRRN} formulated SR as a recursive learning task. The use of recursive residual blocks vastly reduces the model parameters, enabling fast and efficient training of substantially deeper CNN models. Lim \etal \cite{EDSR} combined the ideas of multi-scale reconstruction and residual learning to achieve superior $HR$ reconstruction at high upscaling factors.
% More recently, Zhang \etal \cite{zhang2019image} proposed an adaptive texture transfer that enriches the output images with target's $HR$ details by conforming to their textual similarities in latent feature space.

\subsection{Perceptual Quality}

Despite vast advances in the architecture design of CNNs, the use of point-estimate loss functions (e.g., mean squared error) consistently gave rise to blurry images \cite{DRCN}. This is because point-estimate loss functions suffer from regression-to-the-mean problem at high upscaling factors. In other words, an optimal point-wise estimator returns the mean of many valid $HR$ interpolations, resulting in blurry $HR$ images. Another line of work that has attracted a lot of attention is the design of objective functions that focus on high-level image semantics over pixel-level details. The path taken by these approaches broadly falls into two classes: (i) directly emphasizing high-level feature reconstruction by optimizing in the latent feature-space of a pre-trained network (i.e., perceptual loss  \cite{JohnsonPerceptual}), (ii) iteratively pushing the distribution of generated SR images closer to the distribution over natural $HR$ images using a discriminator network (i.e., adversarial training \cite{GAN_Goodfellow, SRGAN}).

Johnson \etal \cite{JohnsonPerceptual} was the first to introduce a perceptual loss in SR, by using the high-level features of an ImageNet \cite{imagenet} trained VGG network \cite{VGGnet} to obtain sharp, visually pleasing images. The problem of SR was also explored in the context of adversarial learning by Ledig \etal \cite{SRGAN}. Further, this approach also uses perceptual loss for efficient reconstruction of finer $HR$ details. Taking inspiration from Gatys \etal \cite{neuralstyle}, Sajjadi \etal \cite{Enhancenet} proposed a texture-matching loss in addition to adversarial and perceptual losses for the reconstruction of high-level texture details in $SR$ images. These approaches further show that the PSNR metric used to measure the reconstruction fidelity in SISR tasks correlates poorly with the human perception of image quality. In their experiments, a network trained using \textit{MSE} achieves a high PSNR score but fails to generate visually pleasing outputs relative to a network trained on perceptual loss or adversarial loss. To address this issue, Wang \etal \cite{esrgan} proposed a 2-stage training framework, a PSNR-oriented training followed by GAN-based fine-tuning, for trading-off fidelity with perceptual quality.

\begin{figure*}[!htb]
    \centering
    \includegraphics[width=0.65\textwidth]{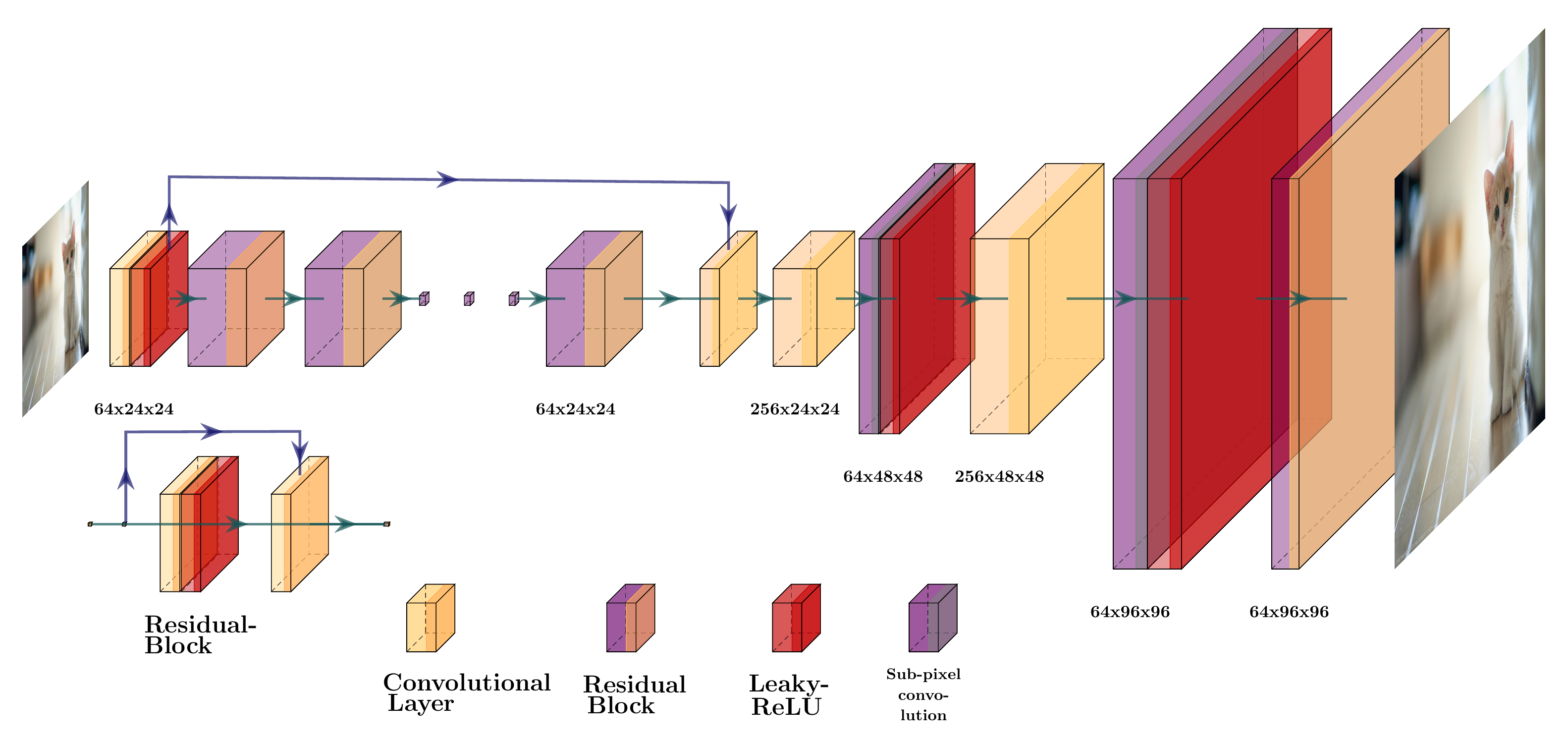}
    \includegraphics[width=0.65\textwidth]{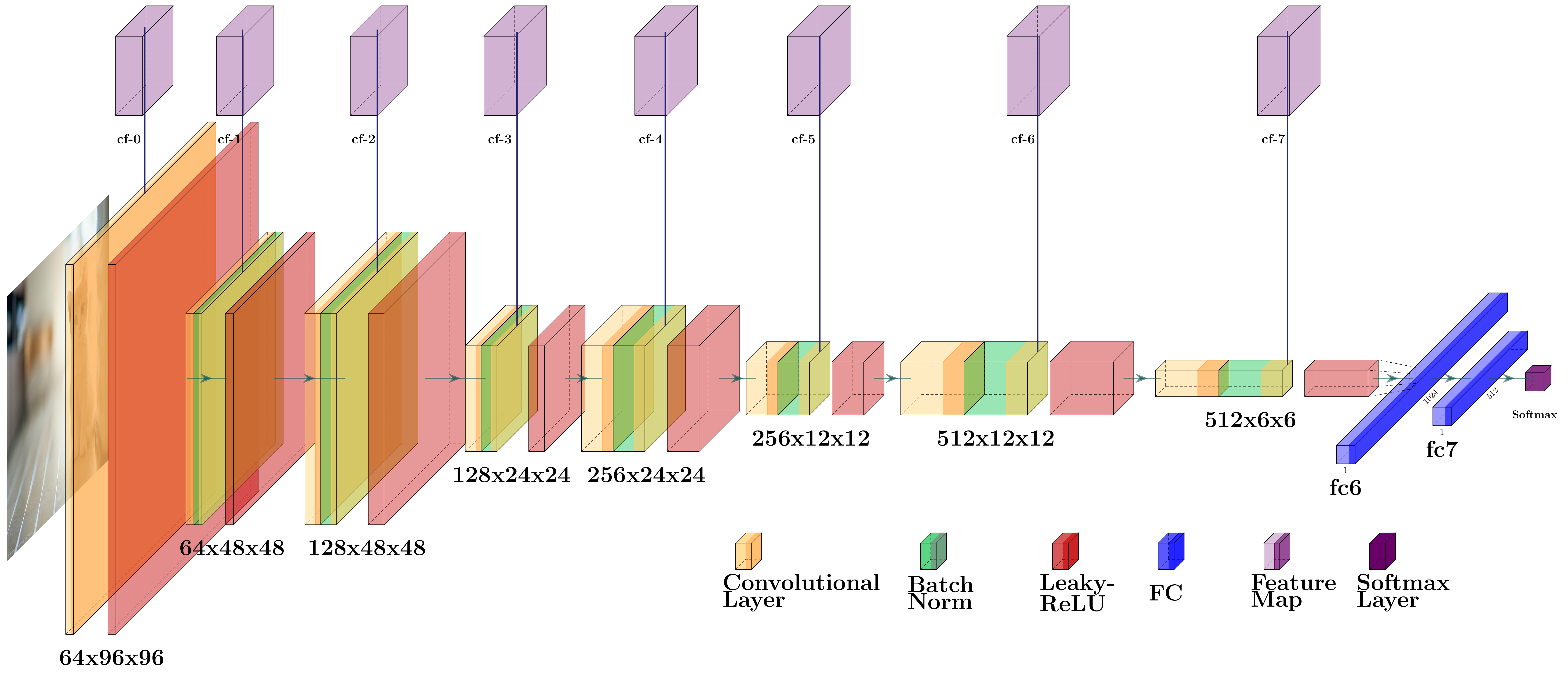}
    \vspace{1em}
    \caption{\textit{Top}: Architecture of our generator network. \textit{Bottom:} Discriminator architecture with feature maps from the $Conv$ layers (in purple) that are used in the computation of our content loss.}
    \label{fig:gen_arch}
\end{figure*}

\section{Proposed Method}

% As the name suggests, these loss functions have demonstrated a positive correlation with the human perception of image quality.

% Since the features of a discriminator are adaptively trained to differentiate between the generated and target $HR$ images, a content loss derived from these features does not implant artifacts that make the generated outputs distinguishable from natural images. Thus, the proposed content loss is an unbiased estimator for the true SR objective.

Perceptual loss functions have several properties that make them appealing in the context of SR, \textit{viz.} (i) they do not suffer from regression-to-the-mean problem like point-estimate loss functions, (ii) the CNN-based architecture of the pre-trained network makes them more stable to local deformations in the $LR$ image, and (iii) they demonstrate a lower variance for stationary textures in the input, which are abundant in natural images. In other words, perceptual loss functions are low variance estimators that can produce stable high-frequency content and, consequently, sharp output images. However, perceptual loss functions were originally trained for a classification task on the ImageNet dataset \cite{imagenet}, which makes them sensitive to the differentiating texture patterns observed across the $1000$ training classes. Thus, using a pre-trained perceptual loss to optimize an SR model often causes unwanted texture patterns in the generated images. 

Our main aim is to efficiently filter out the artifacts introduced by the pre-trained perceptual loss function. We accomplish this by introducing a new content loss formulation that extends the traditional adversarial training framework. Unlike prior SR approaches that only use the final layer of a discriminator network, we derive our content loss from all its latent $Conv$ layers. As a result, our approach provides stronger supervision for the generator's training while also stabilizing non-convex optimization in GANs (more details can be found in Sec. \ref{sec:related_prev_work}). Unlike perceptual loss, the proposed content loss matches the true SR objective for the following reasons:
\begin{itemize}
\itemsep0em
\item[$\bullet$] The training data for the discriminator, i.e., the proposed content loss network, is sampled from the distribution implicitly modeled by the generator network itself. In contrast, perceptual loss \cite{JohnsonPerceptual} network was originally trained for a discriminative task on the ImageNet dataset.
\item[$\bullet$] The discriminator network adaptively learns features that are most discriminative of generated SR outputs versus natural $HR$ images. Thus, the feature space of a discriminator network offers a natural choice of statistics for the generator to match. Crudely speaking, the proposed content loss advocates photo-realism, a sub-goal of SR.
\end{itemize}

\subsection{Network Architecture}
We use the SRGAN \cite{SRGAN} architecture in all our experiments, to have a fair comparison with prior works.

\textbf{Generator: }A fully-convolutional feed-forward network comprising of an encoder and a decoder module. The encoder consists of a stack of $N = 16$ identical residual blocks. Each residual block consists of two $Conv$ layers with $3 \times 3$ kernels, $64$ channels and a LeakyReLU $(\alpha=0.2)$ activation function. The up-sampling decoder consists of two sub-pixel convolutional layers \cite{subpixelCNN}, each increasing the resolution by a factor of $2\times$. In contrast to the original SRGAN architecture, we avoid using Batch-Normalization layers in the generator because of its insensitivity to changes in input statistics, leading to unwanted artifacts and limited generalizability \cite{esrgan}.

\textbf{Discriminator: }This network comprises of $8$ $Conv$ layers using a $3 \times 3$ kernel of stride length $1$ and $4 \times 4$ kernel with stride length $2$ in an alternating fashion. The number of channels in these $Conv$ layers increases linearly from $64$ to $512$ as we go deeper into the architecture. There exists a Batch-Normalization layer and a LeakyReLU $(\alpha=0.2)$ activation between every two $Conv$ layers. The last $Conv$ layer is followed by $2$ $Dense$ layers and a $Sigmoid$ neuron that outputs the final probability.
The complete architecture is displayed in Fig. \ref{fig:gen_arch}.
% The objective of this network is to distinguish the true $HR$ images from the generated SR images by estimating a probability of whether the image is real.

\subsection{Objective Function}

We formulate the overall SR objective ($\Lagr$) as a weighted combination of the following loss functions:

\begin{equation}
    \label{eq:total_loss}
    \begin{aligned}
        \Lagr = \Lagr_{content} + \lambda\Lagr_{adv} + \eta\Lagr_{point} + \gamma\Lagr_{vgg}
    \end{aligned}
\end{equation}

\textbf{Point-estimate loss} focuses on the reconstruction of low frequency components in the generated SR images. Unlike previous approaches that exclusively use $L1$ \cite{EDSR} or $L2$ loss \cite{VDSR, Enhancenet} for this purpose, we use Huber loss  \cite{huber}, a hybrid of $L1$ and $L2$ losses. Huber loss provides a robust loss function for regression that is less sensitive to outliers than $L2$ loss and more stable than $L1$ loss. Huber loss is defined as
\begin{equation}
    \label{eq:huber_loss}
    \Lagr_{point} = \begin{cases}
    \frac{1}{2}||I_{est} - I_{HR}||^2, & \text{if $|I_{est} - I_{HR}| < 1$}.\\
    |I_{est} - I_{HR}| - 0.5, & \text{otherwise},
  \end{cases}
\end{equation}
\noindent where the low-resolution images, estimated super-resolution images and target high-resolution images are denoted by $I_{LR}$, $I_{est}$ and $I_{HR}$ respectively. Both the $L1$ and $L2$ losses in Eq. \ref{eq:huber_loss} also include averaging over all the image dimensions, which is not explicitly written for simplicity.

\textbf{Perceptual loss} \cite{neuralstyle} computes the squared $L2$ norm between the target $HR$ and the output SR images in the latent feature space of a pre-trained VGG network \cite{VGGnet}. Optimizing with perceptual loss instead of pixel-wise losses constrains the generator to produce images that match the high-level feature representations of the target images, thereby enforcing the reconstruction of high-frequency components in $HR$ space. Let $\psi_i$ denote the $i^{th}$ feature layer of the $VGG19$ network. $\Lagr_{vgg}$ is defined as:

\begin{equation}
    \begin{aligned}
    % \sum_{i}\frac{1}{C_iH_iW_i}
        \Lagr_{vgg} = ||\psi_i(I_{est}) - \psi_i(I_{HR})||^2
    \end{aligned}
\end{equation}

\textbf{Adversarial loss} \cite{GAN_Goodfellow} directly optimizes for photo-realism. The adversarial framework involves the joint-training of two networks, a generator $G$ and a discriminator $D$. The generator loss is defined as the negative log-probability of discriminator for the generator's outputs.
\begin{equation}
    \begin{aligned}
        \label{eq:adv_loss}
        \Lagr_{adv} = \mathop{\mathbb{E}_{I_{LR}}} [-\log(D(G(I_{LR})))]
        %\Lagr_{adv} = min_G max_D \mathop{\mathbb{E_{I_{LR}}}} [\log(1 - D(G(I_{LR})))] + \\ \mathop{\mathbb{E_{I_{LR},I_{HR}}}}[\log D(I_{HR})]
    \end{aligned}
\end{equation}
$D$ is optimized over an opposing objective $\Lagr_{D}$ to differentiate the generated images $I_{est}$ from the target images $I_{HR}$.
\begin{equation}
    \label{eq:discriminator_loss}
    \resizebox{0.9\linewidth}{!}{
        $\Lagr_{D} = \mathop{\mathbb{E}_{I_{LR}}} [-\log(1 - D(G(I_{LR})))] + \mathop{\mathbb{E}_{I_{HR}}}[-\log D(I_{HR})]$
    }
\end{equation}

\textbf{Content loss} extends the standard adversarial loss as the squared $L2$ norm across all the latent $Conv$ feature maps of the discriminator network $D$ for $I_{HR}$ and $I_{est}$ images. Since the layers in $D$ learn a hierarchy of differentiating representations of real and fake images, we optimize over an ensemble of content losses derived from all the $Conv$ feature maps. We do not consider dense features for this purpose as they lose spatial information, limiting their utility in the reconstruction of high-frequency components. Further, we use pre-activated features for computing the content loss as the activation layer sparsifies the feature maps and consequently weakens discriminator supervision \cite{esrgan}.

Let $\phi_{i}$ denote a function that returns the pre-activated feature maps corresponding to the $i^{th}$ $Conv$ block of $D$. Then, the $i^{th}$ content loss is defined as
\begin{equation}
    \begin{aligned}
        \label{eq:content_singlelayer}
        % \frac{1}{C_{i}H_{i}W_{i}}
        \Lagr^{i}_{content} = ||\phi_{i}(I_{est}) - \phi_{i}(I_{HR})||^2
    \end{aligned}
\end{equation}

Since there generally exist several $Conv$ layers in $D$ with each layer learning a unique abstraction for differentiating between the natural $HR$ and generated SR images, we wish to optimize over all the content losses simultaneously. While simply averaging over all the layer-wise content losses provides satisfactory results, such a scheme would not equitably optimize over all the content losses. This is because different content losses have varying optimization landscapes and a fixed weighted-averaging scheme would often provide an overall gradient that disproportionately favors only a subset of content losses. Moreover, selecting the weight of each loss over the course of training is also a non-trivial task.

To address this issue, we propose \textbf{softmax reweighing}, a dynamic mechanism to select the weight of each layer-wise content loss such that they are equitably optimized. Before starting the training, we re-weight the individual content losses to bring them to a comparable scale. During training, we rescale the gradient of each content loss by the softmax average of the content loss itself. Thus, during any update, the softmax averaging favors the optimization of a content loss with greater value over one with a smaller value. The overall content loss $\Lagr_{content}$ is formulated as,

\begin{equation}
    \begin{aligned}
        \label{eq:content_total}
        \nabla_{\theta}\Lagr_{content} &= \sum_{i}\bigg\{\frac{e^{\Lagr^{i}_{content}}}{\sum_{j}e^{\Lagr^{j}_{content}}}\bigg\} \nabla_{\theta}\Lagr^{i}_{content}\\
        \Lagr_{content} &= \sum_{i}\bigg\{\frac{e^{\Lagr^{i}_{content}}}{\sum_{j}e^{\Lagr^{j}_{content}}}\bigg\}\Lagr^{i}_{content}
    \end{aligned}
\end{equation}
\noindent where $\theta$ are the parameters of $G$ and $\{.\}$ prevents gradient back-propagation (we do not compute the gradient for the softmax operation). In practice, we found this trick to evenly optimize over all the content losses and subsequently provide a notable improvement in the generator's performance.

\section{Connections with Prior Work}
\label{sec:related_prev_work}

Since perceptual loss functions are derived from the latent features of a pre-trained network, a lot of its properties can be traced back to its pre-training strategies and the ImageNet dataset. More specifically,  Geirhos \etal \cite{Imagenet_CNN_bias_texture} showed that the features extracted by ImageNet-trained CNNs are sensitive to the texture patterns in the images. Complementary to our analysis, they investigate training strategies to obtain better feature extractors that are more robust to texture patterns and better match the human perception of image quality.

Using the intermediate layers of a discriminator for deriving a content loss function improves the stability of adversarial training \cite{GAN_unstable_2}. This prevents the generator from over-training on the output statistics of a discriminator, while also encouraging it to model the multi-modal distribution of natural $HR$ images. As a result, our content loss formulation provides the generator with greater supervision from the discriminator, and subsequently encourages the convergence of GANs, i.e., finding the Nash equilibrium of the minmax game.

\section{Experiments}
\label{sec:experiments}

\subsection{Training Details}
% \label{sub:exp_settings}
\textbf{Data Preparation:} Our training data consists of 800 high-quality images from the DIV2K \cite{div2k} train set and 2650 high-quality images from the Flickr2K \cite{flickr2k} dataset. All the SR experiments are conducted for a $4\times$ scale factor between $LR$ and $HR$ images, i.e., $16\times$ increase in image pixels. We extract random $LR$ patches of spatial dimension $24\times 24$, which correspond to $HR$ patches of size $96 \times 96$. The initial training data is augmented with $90^\circ$ rotations, horizontal and vertical flips. We observe that making the input data zero-centered (i.e., subtracting with the mean of the entire training dataset) provides an improvement in the generator's performance. For testing, we use 4 standard benchmark datasets: Set5 \cite{set5}, Set14 \cite{set14}, BSD100 \cite{bsd100} and Urban100 \cite{selfexsr}. 

\textbf{Training Parameters:} We choose the MSRA initialization \cite{msra_initialization} for the weights of our network but further multiply them by 0.1, as reduced variance in the initial weights helps with faster convergence. Initial learning rate is set to 1e-4 for both the networks and reduced by a factor of 10 after every 200 epochs. Adam optimizer \cite{kingma2014adam} is used to update the generator and discriminator networks. The weights $(\lambda, \eta, \gamma)$ of $\Lagr_{adv}, \Lagr_{point}$ and $\Lagr_{vgg}$ are set to 0.005, 0.01 and 0.5 respectively. All the SR models are trained for $2\times10^5$ updates with a batch size of 16.
% We use PyTorch as our deep learning library and train our networks for 36 hours on a Nvidia GeForce GTX 1080 Ti.

\subsection{Experimental Design}
We train our SR network with a few variations in the overall loss function to systematically investigate the effect of different loss components on the generated outputs.
% More specifically, we wish to study the contribution of the proposed content loss and softmax reweighing.
We investigate the following combinations of loss components with their corresponding trained SR models:
% We perform an extensive quantitative and qualitative ablation study with the following combinations of loss components with their corresponding trained SR models
% We perform an extensive quantitative as well as qualitative ablation study using these models:
% We train our SR network using different combination of loss functions to investigate the effect of different loss components on the perceptual quality and fidelity of the generated outputs.
% The ablation study consists of 5 different combinations of loss functions and their corresponding trained SR models as mentioned below.
\begin{itemize}
    \item[$\bullet$] $\Lagr_{point} \xrightarrow{} M_p$
    \item[$\bullet$] $\Lagr_{point}+\Lagr_{vgg} + \Lagr_{adv} \xrightarrow{} M_{pva}$ 
    \item[$\bullet$] $\Lagr_{point} + \Lagr_{content} + \Lagr_{adv} \xrightarrow{} M_{pca}$
    \item[$\bullet$] $\Lagr_{point} + \Lagr_{content}(softmax) + \Lagr_{adv} \xrightarrow{} M_{pc\sigma a}$
    \item[$\bullet$] $\Lagr_{point} + \Lagr_{content}(softmax) + \Lagr_{vgg} + \Lagr_{adv} \xrightarrow{} M_{pc\sigma va}$
\end{itemize}
Further, we also compare the reconstruction fidelity and perceptual quality of our final model with other SR models.
% both quantitatively and qualitatively.
%\textbf{Comparisons with State-of-the-Art SR Methods:} We compare the performance of our final SR model with other popular SR models, both quantitatively and qualitatively. In this experiment, the performance of an SR model is measured in terms of reconstruction fidelity and perceptual quality.
% \subsubsection{Mean Opinion Score}
% For Mean Opinion Score results, we calibrate the raters on Nearest-Neighbour (score 1) and $HR$ (score 5). Each rater rated 8 versions of 20 randomly sampled images from BSD100: NN, bicubic, $M_{p}$, $M_{pva}$, $M_{pca}$, $M_{pc\sigma a}$, $M_{pc\sigma va}$, and $HR$.

\subsection{Quality Metrics}
% \subsection{Human Evaluation Study}
% We perform a quantitative analysis for the reconstruction fidelity of the proposed method with other comparative SR models.
\textbf{Mean Opinion Score (MOS):} This metric assesses the perceptual quality of the generated images by gathering opinion scores from human raters. We use this metric to compare SR models trained with different combinations of loss components, thereby examining the influence each loss component on the perceptual quality. For the MOS test, we asked $25$ human raters to assess the perceptual quality of the images with an integral score of $1$ (low perceptual quality) to $5$ (high perceptual quality). The human raters were first calibrated with $5$ examples of Nearest Neighbor (NN) (score: 1) and $HR$ images (score: 5). Subsequently, each human rater was given $8$ versions of $20$ randomly-presented images from BSD100: NN, bicubic, $M_{p}$, $M_{pva}$, $M_{pca}$, $M_{pc\sigma a}$, $M_{pc\sigma va}$, and $HR$ (ground-truth). In other words, each of the $100$ images (including its 8 versions) from the BSD100 dataset received a score from $5$ human raters.
\begin{figure}[!htbp]
    \centering
    \includegraphics[width=0.9\linewidth]{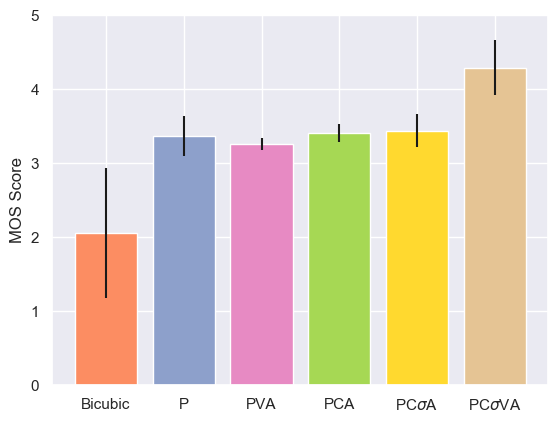}
    \caption{The MOS scores for our model trained with different combinations of loss components.}
    \label{fig:MOS_lossfunctions}
\end{figure}

\textbf{Objective Evaluation:} We report the peak signal-to-noise-ratio (PSNR), a standard evaluation metric in super-resolution for measuring the content preservation, i.e., reconstruction fidelity. Since PSNR correlates poorly with the human perception of image quality, we also report the performance on structural similarity (SSIM), and visual information fidelity (VIF) metrics.
% for outputs of the proposed method versus state-of-the-art SR methods, we use the following standard automatic metrics: peak signal-to-noise-ratio (PSNR), structural similarity (SSIM) and visual information fidelity (VIF). While PSNR is the standard benchmark for SR tasks, we additionally also report SSIM and VIF scores as they better correlate with the human perception of image quality.
%  We included SSIM and VIF scores as additional comparisons because PSNR scores correlate poorly with the human perception of image quality. 
More specifically, SSIM uses spatial correlation, contrast distortion, and luminance masking to estimate the image quality. On the other hand, Visual Information Fidelity (VIF) uses natural scene statistics (NSS) in addition to an image degradation system and a human visual system (HVS) model for image assessment.

%%%Table:

\begin{table*}[!htbp]
\centering
\caption{Quantitative Comparison of PSNR/SSIM/VIF values on test datasets for different combinations of loss components.} 

\label{tb:quant_baseline}
\resizebox{\linewidth}{!}{%
\begin{tabular}{ccccccc}
%  & \multicolumn{5}{c}{\textit{PSNR/SSIM/CIEDE2000}}\\
\hline
\multicolumn{1}{c}{\textit{Dataset}} & \multicolumn{1}{c}{\textit{Bi-cubic}} & \multicolumn{1}{c}{$M_p$} & \multicolumn{1}{c}{$M_{pva}$} & \multicolumn{1}{c}{$M_{pca}$} & \multicolumn{1}{c}{$M_{pc\sigma a}$} & \multicolumn{1}{c}{$M_{pc\sigma va}$}\\ \hline
Set5 & 28.42/0.810/0.443 & \textcolor{teal}{\textbf{31.77/0.890/0.575}} & 23.49/0.848/0.505 & \textcolor{magenta}{\textbf{30.69/0.875/0.529}} & \textcolor{blue}{\textbf{30.76/0.882/0.554}} & 30.03/0.864/0.520\\
Set14 & 26.00/0.704/0.380 & \textcolor{teal}{\textbf{28.40/0.778/0.472}} &	21.88/0.725/0.403 &	\textcolor{magenta}{\textbf{27.47/0.76/0.425}} & \textcolor{blue}{\textbf{27.55/0.761/0.446}} & 26.74/0.742/0.418\\
BSD100 \cite{bsd100} & 25.96/0.669/0.364 & \textcolor{teal}{\textbf{27.46/0.732/0.422}} & 21.81/0.674/0.347 & \textcolor{magenta}{\textbf{26.86/0.714/0.390}} & \textcolor{blue}{\textbf{26.72/0.715/0.402}}	& 26.17/0.693/0.372\\
Urban100 \cite{selfexsr} & 23.15/0.659/0.371 & \textcolor{teal}{\textbf{25.78/0.776/0.446}}	& 20.82/0.715/0.371 & \textcolor{magenta}{\textbf{24.80/0.754/0.393}} & \textcolor{blue}{\textbf{24.97/0.765/0.412}} & 24.40/0.744/0.380 \\\hline
\end{tabular}%
}
\end{table*}

\begin{table*}[!htbp]
\centering
\caption{Peak Signal-to-Noise-Ratio (PSNR) values of different SR methods on test datasets.}
\label{tb:quant_PSNR}
\resizebox{\linewidth}{!}{%
\begin{tabular}{ccccccccccccccc}
%  & \multicolumn{5}{c}{\textit{PSNR/SSIM/CIEDE2000}}\\
\hline
\multicolumn{1}{c}{\textit{Dataset}} &
\multicolumn{1}{c}{Bi-cubic} & 
\multicolumn{1}{c}{DRCN} & 
\multicolumn{1}{c}{DRRN} &
\multicolumn{1}{c}{DSRN} & 
\multicolumn{1}{c}{EDSR} &
\multicolumn{1}{c}{E-Net} &
\multicolumn{1}{c}{ESRGAN}&
\multicolumn{1}{c}{LapSRN}&
\multicolumn{1}{c}{SelfExSR} & 
\multicolumn{1}{c}{SRCNN} & 
\multicolumn{1}{c}{SRGAN} & 
\multicolumn{1}{c}{VDSR} & 
\multicolumn{1}{c}{\textit{Ours}} & 
\multicolumn{1}{c}{\textit{Ours}}\\
& &\cite{DRCN} & \cite{DRRN} & \cite{DSRN} & \cite{EDSR} & \cite{Enhancenet} &\cite{esrgan} & \cite{LapSRN} & \cite{selfexsr} & \cite{SRCNN_1} & \cite{SRGAN}& \cite{VDSR} & $M_{pc\sigma va}$ & $M_{p}$ \\[0.5em]\hline
Set5 & 28.42 & 31.54 & \textcolor{magenta}{\textbf{31.68}} & 31.40 & \textcolor{teal}{\textbf{32.64}} & 28.57 & 30.47 & 31.74 & 30.34 & 30.08 & 29.92 & 31.35 & 30.03 & \textcolor{blue}{\textbf{31.77}}\\

Set14 & 26.00 & 28.12 & \textcolor{magenta}{\textbf{28.21}} & 28.07 & \textcolor{teal}{\textbf{28.95}} & 25.77 & 26.29 & 28.26 & 27.55 & 27.27 & 26.57 & 28.03 & 26.74 & \textcolor{blue}{\textbf{28.40}}\\

BSD100 & 25.96 & 27.23 & \textcolor{magenta}{\textbf{27.38}} & 27.25 & \textcolor{teal}{\textbf{27.80}} & 24.93 & 25.32 & 27.43 & 26.85 & 26.7 & 25.5 & 27.29 & 26.17 & \textcolor{blue}{\textbf{27.46}}\\

Urban100 & 23.15 & 25.13 & 25.44 & 25.08 & \textcolor{teal}{\textbf{26.86}} & 23.54 & 24.32 & \textcolor{magenta}{\textbf{25.51}} & 24.82 & 24.14 & 24.39 & 25.18 & 24.40 & \textcolor{blue}{\textbf{25.78}}\\\hline
\end{tabular}%
}
\end{table*}

\begin{table*}[!htbp]
\centering
\caption{Structural Similarity (SSIM) values of different SR methods on test datasets.}
\label{tb:quant_SSIM}
\resizebox{\linewidth}{!}{%
\begin{tabular}{ccccccccccccccc}
%  & \multicolumn{5}{c}{\textit{PSNR/SSIM/CIEDE2000}}\\
\hline
\multicolumn{1}{c}{\textit{Dataset}} &
\multicolumn{1}{c}{Bi-cubic} & 
\multicolumn{1}{c}{DRCN} & 
\multicolumn{1}{c}{DRRN} &
\multicolumn{1}{c}{DSRN} & 
\multicolumn{1}{c}{EDSR} &
\multicolumn{1}{c}{E-Net} &
\multicolumn{1}{c}{ESRGAN}&
\multicolumn{1}{c}{LapSRN}&
\multicolumn{1}{c}{SelfExSR} & 
\multicolumn{1}{c}{SRCNN} & 
\multicolumn{1}{c}{SRGAN} & 
\multicolumn{1}{c}{VDSR} & 
\multicolumn{1}{c}{\textit{Ours}} & 
\multicolumn{1}{c}{\textit{Ours}}\\

& &\cite{DRCN} & \cite{DRRN} & \cite{DSRN} & \cite{EDSR} & \cite{Enhancenet} &\cite{esrgan} & \cite{LapSRN} & \cite{selfexsr} & \cite{SRCNN_1} & \cite{SRGAN}& \cite{VDSR} & $M_{pc\sigma va}$ & $M_{p}$ \\[0.5em]\hline

Set5 & 0.810 & 0.885 & \textcolor{magenta}{\textbf{0.889}} & 0.883 & \textcolor{teal}{\textbf{0.900}} & 0.81 & 0.852 & \textcolor{magenta}{\textbf{0.889}} & 0.863 & 0.853 & 0.851 & 0.882 & 0.864 & \textcolor{blue}{\textbf{0.890}} \\
Set14 & 0.704 & 0.769 & 0.772 & 0.770 & \textcolor{teal}{\textbf{0.790}} & 0.678 & 0.698 & \textcolor{magenta}{\textbf{0.774}} & 0.755 & 0.743 & 0.709 & 0.770 & 0.742 & \textcolor{blue}{\textbf{0.778}}\\
BSD100 & 0.669 & 0.723 & 0.728 & 0.724 & \textcolor{teal}{\textbf{0.744}} & 0.626 & 0.65 & \textcolor{magenta}{\textbf{0.731}} & 0.711 & 0.702 & 0.652 & 0.726 & 0.693 & \textcolor{blue}{\textbf{0.732}}\\
Urban100 & 0.659 & 0.751 & 0.764 & 0.747 & \textcolor{teal}{\textbf{0.808}} & 0.693 & 0.733 & \textcolor{magenta}{\textbf{0.768}} & 0.739 & 0.705 & 0.731 & 0.753 & 0.744 & \textcolor{blue}{\textbf{0.776}} \\\hline
\end{tabular}%
}
\end{table*}

\begin{table*}[!htbp]
% \label{tb:VIF_1}

\begin{minipage}[b]{0.75\linewidth}
\caption{Visual Information Fidelity (VIF) values of different SR methods on test datasets.}
\resizebox{\textwidth}{!}{%
\begin{tabular}{ccccccccccc}
%  & \multicolumn{5}{c}{\textit{PSNR/SSIM/CIEDE2000}}\\
\hline
\multicolumn{1}{c}{\textit{Dataset}} &
\multicolumn{1}{c}{Bi-cubic} & 
\multicolumn{1}{c}{DRCN} & 
\multicolumn{1}{c}{EDSR} &
\multicolumn{1}{c}{E-Net} &
\multicolumn{1}{c}{ESRGAN}&
\multicolumn{1}{c}{SelfExSR} & 
\multicolumn{1}{c}{SRCNN} & 
\multicolumn{1}{c}{SRGAN} & 
\multicolumn{1}{c}{\textit{Ours}} & 
\multicolumn{1}{c}{\textit{Ours}}\\

& &\cite{DRCN} & \cite{EDSR} & \cite{Enhancenet} &\cite{esrgan} & \cite{selfexsr} & \cite{SRCNN_1} & \cite{SRGAN} & $M_{pc\sigma va}$ & $M_{p}$ \\[0.5em]\hline

Set5 & 0.443 & \textcolor{magenta}{\textbf{0.540}} & \textcolor{blue}{\textbf{0.574}} & 0.44 & 0.502 & 0.502 & 0.48 & 0.49 & 0.520 & \textcolor{teal}{\textbf{0.575}}\\
Set14 & 0.380 & \textcolor{magenta}{\textbf{0.418}} & \textcolor{blue}{\textbf{0.452}} & 0.346 & 0.376 & 0.398 & 0.377 & 0.37 & \textcolor{magenta}{\textbf{0.418}} & \textcolor{teal}{\textbf{0.472}}\\
BSD100 & 0.364 & 0.361 & \textcolor{blue}{\textbf{0.387}} & 0.293 & 0.313 & 0.346 & 0.333 & 0.303 & \textcolor{magenta}{\textbf{0.372}} & \textcolor{teal}{\textbf{0.422}}\\
Urban100 & 0.371 & \textcolor{magenta}{\textbf{0.380}} & \textcolor{teal}{\textbf{0.452}} & 0.332 & 0.377 & 0.365 & 0.325 & 0.36 & \textcolor{magenta}{\textbf{0.380}} & \textcolor{blue}{\textbf{0.446}}
\end{tabular}
\label{tb:VIF_1}
}
\end{minipage}\qquad
\begin{minipage}[b]{0.20\linewidth}
    \resizebox{\textwidth}{!}{%
    \fbox{\begin{tabular}{ll}
        \textcolor{teal}{$\blacksquare$} & Best score \\
        \textcolor{blue}{$\blacksquare$} & $2^{nd}$ best score \\
        \textcolor{magenta}{$\blacksquare$} & $3^{rd}$ best score
    \end{tabular}}
}%
\end{minipage}
\end{table*}

\section{Results and Analysis}
\label{sub:results}

\subsection{Quantitative Analysis}
% \textbf{Ablation Study:}
The results from the MOS test are displayed in Fig. \ref{fig:MOS_lossfunctions}. We observed that the ratings for identical images did not show much variance and a majority of the users rated NN and $HR$ images as $1$ and $5$ respectively. The results in Fig~\ref{fig:MOS_lossfunctions} indicate that $M_{pc\sigma va}$ model 
% generates the most perceptually convincing images, 
significantly outperforms our other models, with an average improvement of over $1$ MOS score. Further, the MOS scores for $M_{pva}$ model are inferior to $M_{pc\sigma va}$, confirming our hypothesis that the proposed content loss brings the best out of the perceptual and adversarial losses. Moreover, removing the perceptual ($M_{pc\sigma a}$) and adversarial loss components ($M_{p}$) results in similar performance degradation, which suggests the complementary nature of these loss components and highlights the importance of their presence for attaining a superior perceptual quality.

Table \ref{tb:quant_baseline} summarizes the performance on objective evaluation metrics (PSNR, SSIM and VIF) for different combinations of loss components. $M_p$ consistently outperforms all our other models since the point estimate loss directly optimizes for reconstruction fidelity. It also explains why $M_p$ falls behind our other models in terms of MOS scores and $M_{pc\sigma va}$ (our model with highest perceptual quality) only delivers a modest performance on objective evaluation metrics.
% $M_{pva}$, $M_{pca}$, $M_{pc\sigma a}$ and $M_{pc\sigma va}$ in terms of MOS scores. Moreover, the $M_{pc\sigma va}$ model is not in the
In summary, none of the objective evaluation metrics from our experiments correlates with the human perception of image quality, i.e., MOS scores. Further, the superior reconstruction fidelity of $M_{pc\sigma a}$ to $M_{pca}$ can be attributed to the softmax reweighing. Another interesting observation is that $M_{pca}$ outperforms $M_{pva}$, suggesting that the proposed content loss (derived from the latent features of a discriminator network) can be a good proxy for the perceptual loss (derived from the latent features of a pre-trained VGG network). 
% $M_p$ is highest across all the automatic metrics but falls behind $M_{pva}$, $M_{pca}$, $M_{pc\sigma a}$ and $M_{pc\sigma va}$ in terms of MOS scores. Surprisingly, $M_{pca}$ outperforms $M_{pva}$, i.e., the discriminator content loss turns out to be a good proxy for the VGG-based perceptual loss. The automatic metrics as well as MOS scores of $M_{pca}$ are behind $M_{pc\sigma a}$, reinforcing the utility of softmax weighting for efficient discriminator content loss optimization.

% \textbf{Comparison with other SR methods:}
The results from Table \ref{tb:quant_PSNR}, \ref{tb:quant_SSIM}, and \ref{tb:VIF_1} suggest that the objective evaluation metrics favor models trained on point estimate loss functions, i.e., EDSR \cite{EDSR} and $M_p$ (trained with Huber loss only). More interestingly, all the perceptually-motivated approaches (e.g., $M_{pc\sigma va}$, SRGAN \cite{SRGAN}, EnhanceNet \cite{Enhancenet}) are outperformed by simple models such as SRCNN \cite{SRCNN_1} trained with point-estimate losses. This supports our previous conclusion that objective evaluation metrics correlate poorly with perceptual quality. Nevertheless, $M_{pc\sigma va}$ still provides a competent reconstruction fidelity relative to other SR methods.

\subsection{Qualitative Analysis}
\label{sec:qualitative}

\begin{figure*}[t]
\sbox{\bigpicturebox}{%
  \begin{subfigure}[b]{.40\textwidth}
  \scalebox{1}[1]{\includegraphics[width=\textwidth]{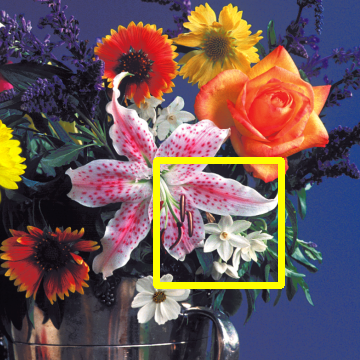}}%
  \caption{HR}
\end{subfigure}
}

\usebox{\bigpicturebox}\hfill
\begin{minipage}[b][\ht\bigpicturebox][s]{.705\textwidth}
\begin{subfigure}{.27\textwidth}
\includegraphics[width=\textwidth]{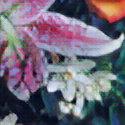}
\caption{Bicubic}
\end{subfigure}
\begin{subfigure}{.27\textwidth}
\includegraphics[width=\textwidth]{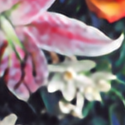}
\caption{$M_p$}
\end{subfigure}
\begin{subfigure}{.27\textwidth}
\includegraphics[width=\textwidth]{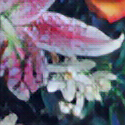}
\caption{$M_{pva}$}
\end{subfigure}\\
\hspace{0.1em}
\begin{subfigure}{.27\textwidth}
\includegraphics[width=\textwidth]{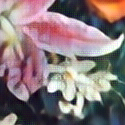}
\caption{$M_{pca}$}
\end{subfigure}
\begin{subfigure}{.27\textwidth}
\includegraphics[width=\textwidth]{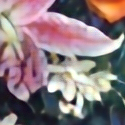}
\caption{$M_{pc\sigma a}$}
\end{subfigure}
\begin{subfigure}{.27\textwidth}
\includegraphics[width=\textwidth]{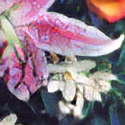}
\caption{$M_{pc\sigma va}$}
\end{subfigure}
\end{minipage}
\vspace{-3.5em}
\caption{Qualitative comparison of outputs from different combinations of loss components on the \textit{Flowers} image from Set14.}
\label{fig:qual_baseline}
\end{figure*}

% \textbf{Ablation Study:}
From Fig. \ref{fig:qual_baseline}, it can be seen that using just the per-pixel loss as in $M_p$ causes the output to blur.
% It seems smoother even when compared to the Bicubic baseline.
The use of perceptual and adversarial losses in $M_{pva}$ results in a sharper image over $M_p$, although the image also tainted with square-like patterns,
% like the bicubic interpolation
giving it with an artificial look. In general, adding perceptual loss increases the sharpness of super-resolved images, as visible from $M_{pva}$ and $M_{pc \sigma va}$. Replacing perceptual loss with the proposed content loss in $M_{pca}$ results in a smoother image with much fewer high-frequency artifacts. The use of softmax reweighing in $M_{pc \sigma a}$ provides a much cleaner image and removes any residual artifacts but still over-smoothens the final output. Finally, $M_{pc \sigma va}$ provides the most perceptually-convincing images with adequate frequency textures and intricate details. Interestingly, removing the perceptual loss causes a noticeable degradation in the perceptual quality for $M_{pc \sigma a}$, emphasizing on the importance of perceptual loss. This qualitative analysis is in coherence with the quantitative analysis of MOS scores.
%where $M_{pc \sigma a}$ is rated as having the highest perceptual quality.

%%%%%%% Figure

%%%% Chinese Baccha %%%%%%%
\begin{figure*}[!htb]
    \centering
    \begin{subfigure}[t]{0.23\linewidth}
    \includegraphics[width=4cm]{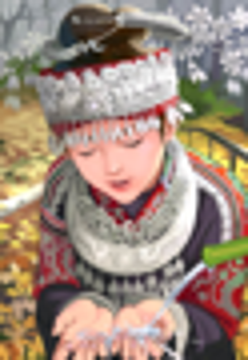}
    \caption{Bicubic}
    \end{subfigure}
    \begin{subfigure}[t]{0.23\linewidth}
    \includegraphics[width=4cm]{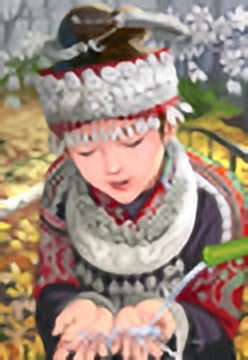}
    \caption{SRCNN \cite{SRCNN_1}}
    \end{subfigure}
    \begin{subfigure}[t]{0.23\linewidth}
    \includegraphics[width=4cm]{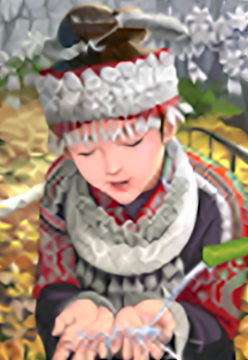}
    \caption{SelfExSR \cite{selfexsr}}
    \end{subfigure}
    \begin{subfigure}[t]{0.23\linewidth}
    \includegraphics[width=4cm]{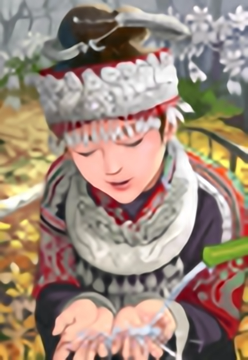}
    \caption{DRCN \cite{DRCN}}
    \end{subfigure} \\
    \begin{subfigure}[t]{0.23\linewidth}
    \includegraphics[width=4cm]{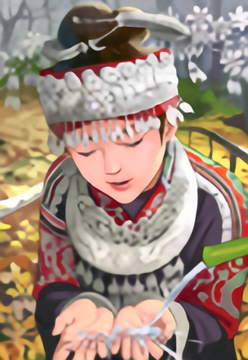}
    \caption{SRResNet \cite{SRGAN}}
    \end{subfigure}
    \begin{subfigure}[t]{0.23\linewidth}
    \includegraphics[width=4cm]{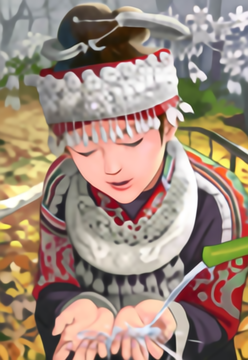}
    \caption{EDSR \cite{EDSR}}
    \end{subfigure}
    \begin{subfigure}[t]{0.23\linewidth}
    \includegraphics[width=4cm]{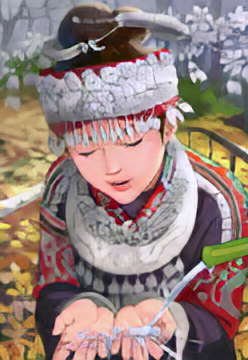}
    \caption{$M_{pc\sigma va}$}
    \end{subfigure}
    \begin{subfigure}[t]{0.23\linewidth}
    \includegraphics[width=4cm]{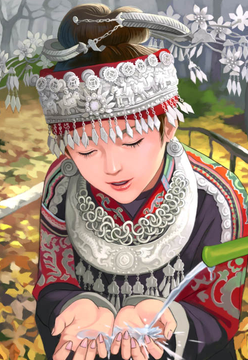}
    \caption{HR}
    \end{subfigure}

\vspace{2em}
\caption{Qualitative comparison of our method ($M_{pc \sigma va}$) with other SR models on the \textit{Comic} image from Set14.}
\label{fig:other_methods}
\end{figure*}

% \textbf{Comparison with other SR methods:}
Fig. \ref{fig:other_methods} displays the outputs of SR methods trained on point-estimate loss functions along with $M_{pc\sigma va}$ in the increasing order of perceptual quality. In contrast to $M_{pc\sigma va}$, the other SR methods lack sharpness and fine details, which can be attributed to regression-to-the-mean problem of point-estimate losses. When compared to perceptual SR methods (see Fig. \ref{fig:first}), $M_{pc\sigma va}$ produces cleaner images with fewer artifacts. More specifically, the building image of SRGAN \cite{SRGAN} contains several box-like artifacts near the edges. The EnhanceNet \cite{Enhancenet} model trained using a VGG-based texture loss has incongruous texture patterns in its super-resolved outputs.
In contrast, our method does not contain any block artifacts or conflicting texture patterns. In other words, although perceptual loss is crucial for obtaining sharp images, we demonstrate that sharpness alone does not directly correlate with the perceptual quality. With increased discriminator supervision, the proposed content loss provides sharp SR images with pertinent high-frequency patterns.

\section{Conclusion}

In this work, we investigated the challenges in training deep generative models with perceptual and adversarial losses for the super-resolution task. We showed that these loss functions have complementary advantages and can be effectively combined to overcome their individual shortcomings. We derived a novel content loss formulation from the latent features of discriminator network to (i) effectively eliminate the biases transferred from the perceptual loss, and (ii) stabilize adversarial training with increased supervision from the discriminator network. Further, we systematically studied the properties of the proposed content loss when combined with other loss functions. Our results confirm that the proposed content loss addresses the individual shortcomings of perceptual and adversarial losses, providing substantial gains in the perceptual quality of the generated images.  Moreover, our approach also provides a competent reconstruction fidelity relative to state-of-the-art SR methods.

\bibliographystyle{IEEEtran}
\bibliography{IEEEabrv,ref}

\end{document}